\documentclass[aps,prb,twocolumn,groupedaddress,nobibnotes,superscriptaddress]{revtex4-1}
\usepackage[english]{babel}
\usepackage{lipsum, babel}
\usepackage{amsmath,amsthm,amssymb,amsfonts} 
\usepackage{graphicx}
\usepackage{xcolor}
\usepackage{color} 
\usepackage{dcolumn}
\usepackage{siunitx}
\usepackage{chngcntr}
\usepackage{hyperref}
\usepackage{mathtools}
\usepackage{graphicx}
\usepackage{float}
\usepackage{placeins}
\usepackage[singlelinecheck=false]{caption}
\usepackage{chemformula}

\begin{document}

\title{
The thermodynamics of  CaSiO\textsubscript{3} in Earth's lower mantle}

\author{Yongjoong Shin}
\affiliation{Theory and Simulation of Materials (THEOS) and National Centre for Computational Design and Discovery of Novel Materials (MARVEL), École Polytechnique Fédérale de Lausanne, Lausanne 1015, Switzerland}

\author{Enrico Di Lucente}
\affiliation{Theory and Simulation of Materials (THEOS) and National Centre for Computational Design and Discovery of Novel Materials (MARVEL), École Polytechnique Fédérale de Lausanne, Lausanne 1015, Switzerland}

\author{Nicola Marzari}
\affiliation{Theory and Simulation of Materials (THEOS) and National Centre for Computational Design and Discovery of Novel Materials (MARVEL), École Polytechnique Fédérale de Lausanne, Lausanne 1015, Switzerland}
\affiliation{Laboratory for Materials Simulations, Paul Scherrer Institut, 5232 Villigen PSI, Switzerland}

\author{Lorenzo Monacelli}
\affiliation{Dipartimento di Fisica, Universit\`a di Roma Sapienza, 00185 Roma, Italy}
\affiliation{Theory and Simulation of Materials (THEOS) and National Centre for Computational Design and Discovery of Novel Materials (MARVEL), École Polytechnique Fédérale de Lausanne, Lausanne 1015, Switzerland}


\begin{abstract}

The lower mantle of Earth, characterized by pressures of 24-127 GPa and temperatures of 1900-2600 K, is still inaccessible to direct observations. In this work, we investigate by first principles the stability, phase diagram, elastic properties, and thermal conductivity of CaSiO$_{3}$, that constitutes a significant component of Earth's lower mantle. Notably, our simulations capture in full the anharmonic ionic fluctuations arising from the extreme temperatures and pressures of the lower mantle, thanks to the use of stochastic self-consistant harmonic approximation (SSCHA).  

We show that the cubic phase of CaSiO$_{3}$ is the stable state at the lower mantle's thermodynamic conditions. The phase boundary between the cubic and tetragonal phases is of first-order and increases linearly from 300~K to 1000~K between \SI{12}{\giga\pascal} and \SI{100}{\giga\pascal}. 
Accounting for temperature-renormalized phonon dispersions, we evaluate the speed of sound as a function of depth. Our results downplay the role of octahedral rotations on the transverse sound velocity of cubic \ch{CaSiO3}, advocated in the past to explain discrepancies between theory and experiments.
The lattice thermal conductivity, assessed thanks to the recently introduced Wigner formalism, shows a predominance of particle-like transport, thus justifying the use of the standard Boltzmann transport equation even in a system with such strong ionic anharmonicity. 



\end{abstract}

\maketitle


Despite our proximity to Earth's lower mantle, the journey to explore it remains far more distant than venturing into the cosmos. The Earth's lower mantle has a temperature around \SI{2000}{\kelvin}\cite{katsura2010adiabatic} and pressures between 24 to \SI{127}{\giga\pascal}\cite{dziewonski1981preliminary}. \ch{CaSiO3} constitutes about 10\% of the lower mantle's mass, and its precise mechanical and thermodynamical characterization under such extreme conditions is still missing. Both experiments and theoretical simulations show that \ch{CaSiO3} exists in the Ca-perovskite structure above \SI{12}{\giga\pascal}\cite{gasparik1994experimental, shen1995measurement, zerr1997melting, braithwaite2019melting}; however, its phase diagram remains mostly unknown. For example, contradictory data have been reported on the crystal symmetry\textemdash cubic (Pm$\overline{3}$m space group) vs tetragonal (I4/mcm space group)\textemdash at high pressures and temperatures\cite{stixrude2007phase, li2006phase, adams2006ab, li2006elasticity}.
Furthermore, seismology of our planet relies on the precise estimation of sound velocities at the relevant thermodynamic conditions\cite{thomson2019seismic}.
However, due to practical challenges in reproducing the high temperature and pressures of lower mantle, most experimental data on the elastic properties of \ch{CaSiO3} have been reported for the tetragonal phase at room temperature\cite{sun2016confirming, kurashina2004phase, ricolleau2010phase, immoor2022weak,miyagi2009diamond, shieh2004elasticity}. Therefore, the propagation speed of longitudinal and transverse seismic waves as a function of depth, necessary for accurate geological modeling of the Earth's interior, remains unknown.
Last, the prediction of lattice thermal conductivities from first principles is a very active area of research, with the complex phenomenology of phonon-phonon scattering in solids being intensively investigated\cite{simoncelli2019unified,phono3py,ShengBTE_2014,carrete2017almabte,raya2022bte,ding2022observation,huberman2019observation,huang2023observation,lee2015hydrodynamic,cepellotti2017boltzmann,sendra2022hydrodynamic,jeong2021transient,restuccia2023non,sykora2023multiscale,di2024vorticity,ponet2024energy}. In the context of geophysics, the knowledge of thermal conductivity is crucial to determine the planet's internal dynamics and evolution, specifically at the core-mantle boundary where mass transport is impeded\cite{Lay2008}.

Computer simulations are thus very precious, as they can reproduce the lower mantle condition \emph{in silico}\cite{alfe1999melting}. However, the strongly anharmonic fluctuations of ions due to high temperature and large quantum fluctuations of lattice phonons pose significant challenges for standard approaches. Previous studies on \ch{CaSiO3} mainly employed the quasi-harmonic approximation (QHA)\cite{stixrude2007phase, sagatova2021phase}, which can describe volume-dependent thermal effects in materials with some degree of anharmonicity. Still, it fails in cubic \ch{CaSiO3}, where harmonic phonons are unstable. Various methods have been employed to overcome this issue\cite{stixrude2007phase, kurashina2004phase, wentzcovitch1995ab, caracas2006theoretical, karki2001high}, the most successful one being first principles molecular dynamics (FPMD) \cite{li2006phase,sokolova2021equations, sun2014dynamic, wu2024deep}, which shows how the crystal remains tetragonal up to 3000 K but appears cubic when averaged over time. This scenario points toward an order-disorder phase transition, for which determining the phase boundary within FPMD simulations is very demanding.

Also, the computational evaluation of the speed of sound and elastic properties \ch{CaSiO3} presents significant challenges.
In fact, the bulk and shear moduli significantly differ between the tetragonal and cubic phases. Thus, it is fundamental to elucidate both the precise phase diagram of the material and the role played by the lattice vibration at several thousands Kelvin\cite{greaux2019sound}. 


The geological modeling of Earth also relies on the determination of heat flows from the Earth's core to the lower mantle. This process is regulated by the thermal conductivity ($\kappa$) of materials composing the core-lower mantle interface\cite{hofmeister2019heat,lodders1998planetary,schubert2015treatise,rost2013core}, of which \ch{CaSiO3} is one of the major constituents.
A recent study reported a systematic calculation of \ch{CaSiO3} thermal conductivity\cite{wentzcovitch2021thermal}. However, coherent wave-like transport\cite{simoncelli2019unified,di2023crossover}
may become relevant at temperatures of several thousands Kelvin, and its role is yet to be investigated.

In this work, we simulate from first principles the phase diagram, sound velocities, and thermal conductivity of \ch{CaSiO3} between 20-\SI{100}{\giga\pascal} and 300-\SI{3000}{\kelvin} accounting for the anharmonic and quantum fluctuations of ions. For this purpose, we combine density-functional theory (DFT) at the PBEsol level\cite{PBEsol}
with the stochastic self-consistent harmonic approximation (SSCHA)\cite{Errea2014a,bianco2017second,Monacelli2018,monacelli2021stochastic,monacelli2024simulating,miotto2024fast}, a method able to compute the thermodynamics properties of solids arising from anharmonic vibrations of the lattice by minimizing the free energy with a Gaussian density matrix ansatz for the ionic degrees of freedom. We calculate the precise phase boundary between the cubic and tetragonal phase of \ch{CaSiO3}, unveiling the origin of the phase transition, and compute its bulk modulus, shear modulus, and thermal conductivity. More details on the computational methodology are provided in the Methods section. 

\begin{figure*}[hbt!]
\centering
\includegraphics[width=\textwidth]{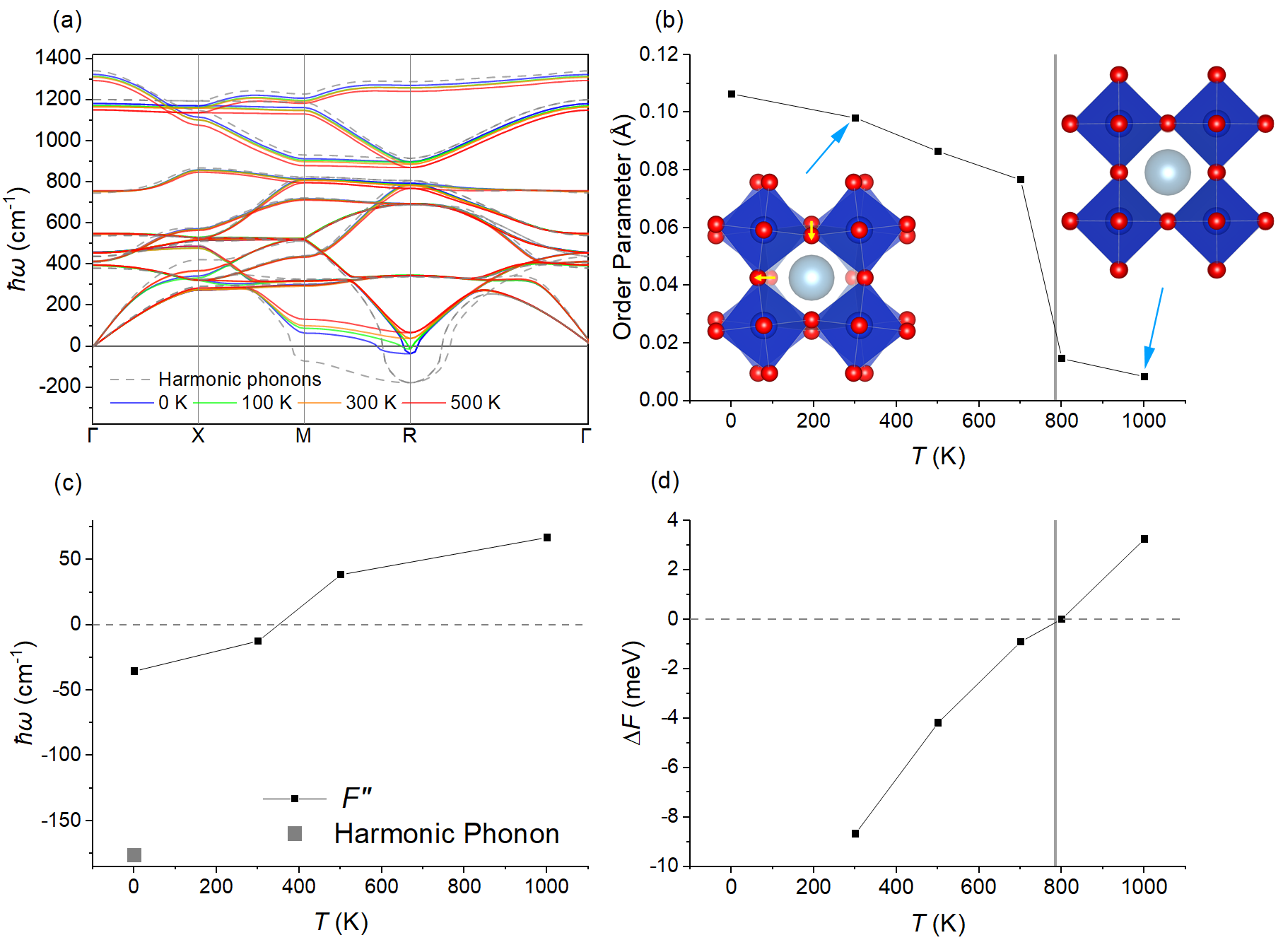}
\caption{Phase stability for \ch{CaSiO3} at \SI{100}{GPa}, across the transition from the cubic Pm$\bar3$m to the tetragonal I4/mcm structure. (a) Anharmonic phonons in cubic \ch{CaSiO3} as a function of temperature (obtained as the square root eigenvalues of the free energy landscape Hessian matrix). Dashed lines indicate the harmonic dispersion computed within density-functional perturbation theory (DFPT). The \SI{0}{\kelvin} dispersions do not match with DFPT, as it also accounts for anharmonicity induced by finite fluctuations of ions due to quantum zero-point motion. 
(b) Order parameter as a function of temperature: the order parameter is chosen as the difference between the average positions of atoms and the Wyckoff positions of the high-symmetry cubic phase. We also show the crystal average structure at 300~K and 1000~K (before and after the transition). (c) The lowest frequency of the free energy Hessian $F'' = \frac{\partial^2F}{\partial R_a\partial R_b}$ at the R point as a function of temperature. Negative (imaginary) values indicate instability. (d) Free energy difference between the cubic and tetragonal phases as a function of temperature; the vertical line identifies the phase transition.}
\label{fig:figure2}
\end{figure*}

\section*{Results}

\subsection{Phase diagram and microscopic origin of the phase transition}
There are contradictory reports on the origin of the phase transition in \ch{CaSiO3}\cite{stixrude2007phase, li2006phase, adams2006ab}. 
In particular,
it is unclear if the transition kind is first-order, order-disorder (similar to \ch{BaTiO3}\cite{pasciak2018dynamic,Gigli2022}), or second-order displacive (as in \ch{CsSnI3}\cite{MonacelliCsSnI32023}).
In the order-disorder scenario, the cubic structure with 5 atoms per primitive cell is unstable even at high temperatures; the system reorganizes its lattice within a larger cell, characterized by disordered local displacements that, on average, retain cubic symmetry\cite{kotiuga2022microscopic}. In the first-order scenario, the cubic and tetragonal structures can coexist in the same thermodynamic conditions: the transition occurs when the free energy of the two phases is the same. The displacive transition is instead a typical second-order phenomenon where the distortion from the cubic lattice (order parameter) continuously disappears with increasing temperature. The displacive transition is marked by a critical point where the correlation function of the order parameter diverges in the thermodynamic limit; this critical point coincides with the softening of a phonon mode, which approaches zero energy.
It is exceptionally challenging for FPMD to disentangle the different mechanisms, as the similarity between the cubic and tetragonal phases prevents a clear identification of the structure from snapshots of the dynamics. Moreover, investigating the existence of the critical point or a disordered cubic structure is cumbersome in finite-size systems, requiring a systematic analysis of the thermodynamic limit through the simulation of cells containing thousands of atoms, often beyond the reach of first-principles calculations. 

To elucidate the mechanism underlying the phase transition, we compare the stability of the 5-atoms cubic phase (\figurename~\ref{fig:figure2}a,c), the evolution of the order parameters in the tetragonal phase (\figurename~\ref{fig:figure2}b), and the free energy differences between cubic and tetragonal structures (\figurename~\ref{fig:figure2}d). We run the simulations at \SI{0}{\giga\pascal}, \SI{50}{\giga\pascal}, and \SI{100}{\giga\pascal} (see also SI. 1 in supplementary information), reporting a detailed analysis in \figurename~\ref{fig:figure2} only for the latter pressure.
The (meta)stability of a structure can be identified from the sign of the anharmonic free energy Hessian eigenvalues\cite{bianco2017second}
 \begin{equation}
     \frac{1}{\sqrt{m_i m_j}}\frac{\partial^2 F}{\partial \mathcal R_{i\alpha} \partial R_{j\beta}} = \sum_\mu {e_\mu}_{i\alpha} {e_\mu}_{j\beta} \omega_\mu^2.
 \end{equation}
Here, $\omega_\mu$ and $\boldsymbol{e_\mu}$ define the $\mu$-th static anharmonic phonon frequency and the respective polarization vector; $m_i$ is the mass of the $i$-th atom; the $i,j$ indicate the atomic indices while $\alpha,\beta$ the Cartesian coordinates. If $\omega_\mu$ is imaginary, the Hessian of the free energy is not positive definite, and the structure is unstable.  
 At \SI{100}{\giga\pascal}, the 5-atoms cubic phase is stable above \SI{300}{\kelvin} (the temperature at which the imaginary phonon at R disappears in \figurename~\ref{fig:figure2}a and c). 
Interestingly, even at \SI{0}{\kelvin}, anharmonic phonon dispersions deviate from the harmonic calculations, emphasizing the importance of anharmonicity even when ions fluctuate only due to quantum zero-point motion, included within the SSCHA\cite{monacelli2021stochastic}.
Increasing the temperature, the rotations of the oxygen octahedra energy (the M-R unstable modes) stiffen. Instead, high-energy phonon bands, representing the expansion and contraction of the oxygen cage, become softer upon heating. Since a temperature above \SI{1000}{\kelvin} is necessary to populate high-energy modes, their redshift at lower temperatures indicates their coupling with a low-energy phonon band. By selectively inhibiting scattering with low-frequency modes in the calculation of the free energy Hessian (see Methods section), we determine that the significant phonon-phonon interactions of the high-energy bands occurs with the rattling of the calcium atoms. On the opposite, the rotational modes of oxygen octahedra that trigger the phase transition do not interact with high energy bands. 

\begin{figure}[hbt!]
\centering
\includegraphics[width=\columnwidth]{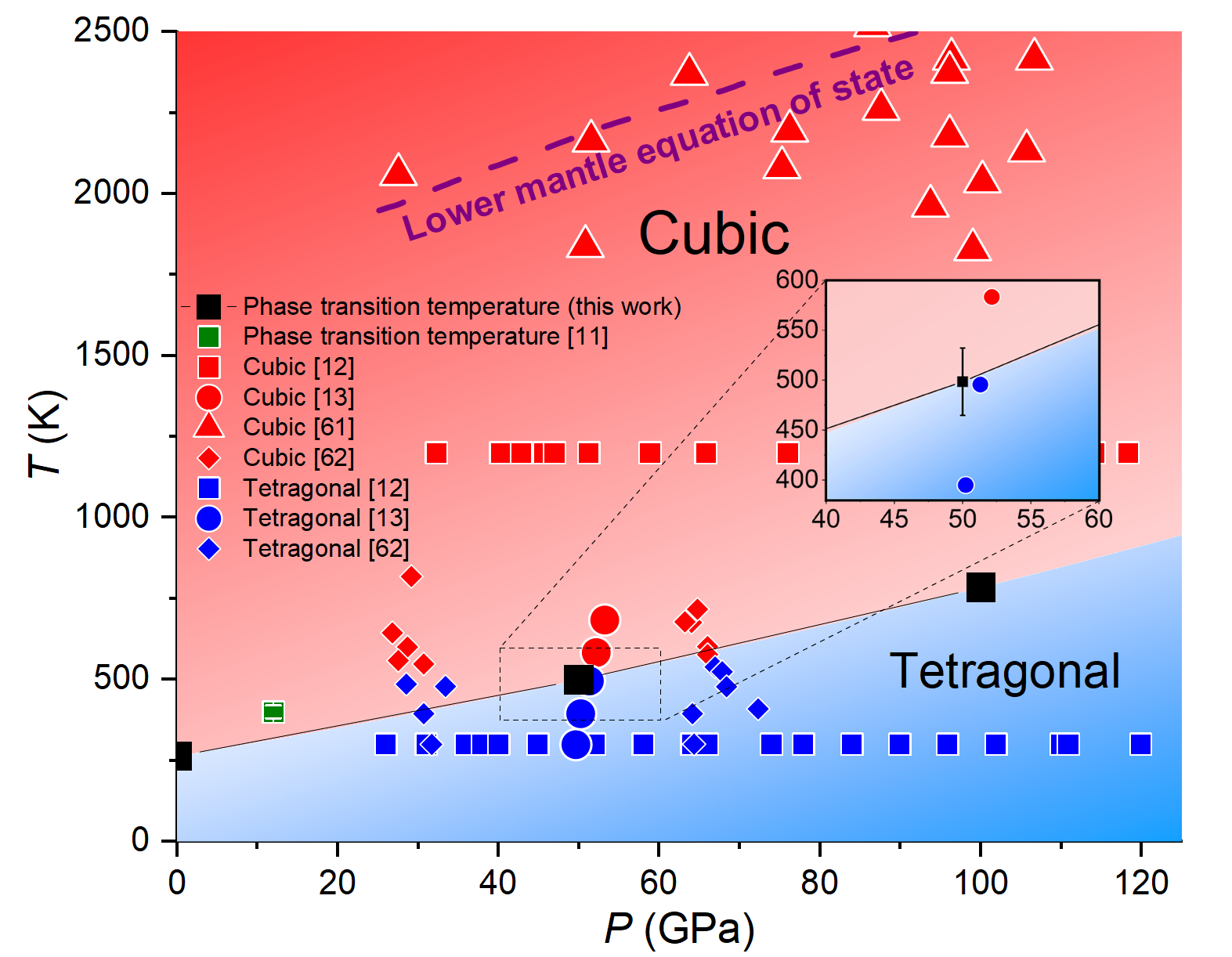}
\caption{Pressure-temperature phase diagram of CaSiO\textsubscript{3}. The phase boundary evaluated in this work is reported with the black line. The error bar has been estimated from the stochastic sampling of the free energy at the transition temperature to be $\sim\SI{30}{\kelvin}$. Experimental data\cite{kurashina2004phase, ono2004phase,sun2016confirming,komabayashi2007phase, thomson2019seismic}, are reported in red (blue) for the cubic (tetragonal) phase. The purple dashed line represents the lower mantle's equation of state.}
\label{fig:Phase Diagram}
\end{figure}

While the simulation of the free energy Hessian shows that a temperature of \SI{400}{\kelvin} is enough to remove the instabilities of the cubic phase at \SI{100}{\giga\pascal}, the free energy of the tetragonal phase remains lower than the cubic one until \SI{800}{\kelvin} (\figurename~\ref{fig:figure2}d). This rules out the second-order displacive phase transition hypothesis: the tetragonal phase has a lower free energy at the critical point where the cubic phase stabilizes. 
In \figurename~\ref{fig:figure2}b we report the order paramete, defined as the difference between the average positions of atoms and the Wyckoff positions of the Pm$\bar3$m cubic phase, as a function of temperature. An abrupt transition into the cubic phase occurs at \SI{800}{\kelvin}. This is consistent with a first-order scenario, where the cubic and tetragonal phases can coexist in a wide range of temperatures between \SI{400}{\kelvin} and \SI{800}{\kelvin}.  
Thanks to the determination of the free energy difference (\figurename~\ref{fig:figure2}d), we identify the transition temperature at \SI{800}{\kelvin} (\SI{100}{\giga\pascal}). 
Taking into account also our simulations at \SI{0}{\giga\pascal} and \SI{50}{\giga\pascal}, we show the resulting phase diagram of \ch{CaSiO3} in \figurename~\ref{fig:Phase Diagram}.
Our calculations agree perfectly with available experimental elastic X-ray diffraction data\cite{kurashina2004phase} and underline how the cubic Pm$\bar3$m phase is stable already at relatively low temperatures and is the dominant phase of \ch{CaSiO3} in Earth's lower mantle. The phase boundary agrees with the recently proposed one based on thermodynamic integration using molecular dynamics with deep learning interatomic potentials\cite{wu2024deep}.

\begin{figure*}[hbt!]
\centering
\includegraphics[width=\textwidth]{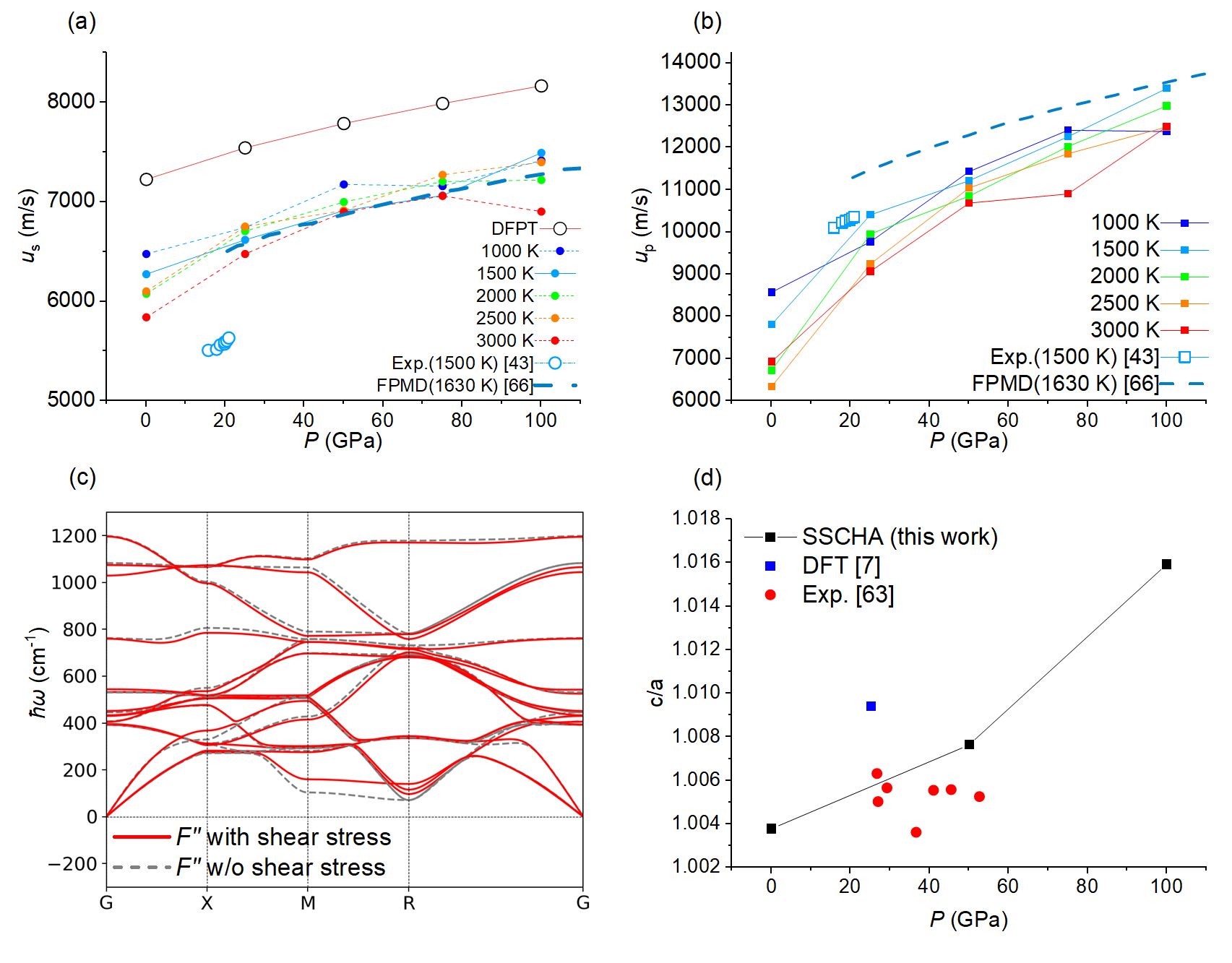}
\caption{(a) Transverse sound velocity $u_s$ and (b) longitudinal sound velocity $u_p$ of \ch{CaSiO3} at different pressures and temperatures. (c) The free energy Hessian $F''=\frac{\partial^2 F}{\partial R_a \partial R_b}$ of \ch{CaSiO3} under shear (at 100 GPa of surrounding pressure, \SI{3000}{\kelvin} and shear of 0.01). The apparent presence of more phonon bands in the deformed structure is due to loss of degeneracy upon cubic symmetry breaking under shear. (d) Lattice parameters ratio (ratio of length of unit cell edges $c$ and $a$) of the I4/mcm unit cell at \SI{300}{\kelvin}.
}
\label{fig:figure3}
\end{figure*}

We report in \figurename~\ref{fig:figure3}d the effects of temperature and pressure on the anisotropy of the tetragonal phase. Here, quantum fluctuations play a non-negligible role that cannot be accounted for with MD or static DFT calculations. In contrast with previous reports\cite{stixrude2007phase}, our simulations of the $c/a$ tetragonal anisotropy show an excellent agreement with the X-ray diffraction data obtained in diamond-anvil cells\cite{chen2018crystal}, and highlight how pressure favors the tetragonal anisotropy, reflected in the positive slope of the phase-boundary with the cubic phase (\figurename~\ref{fig:Phase Diagram}a).

\begin{figure*}[hbt!]
\centering
\includegraphics[width=\textwidth]{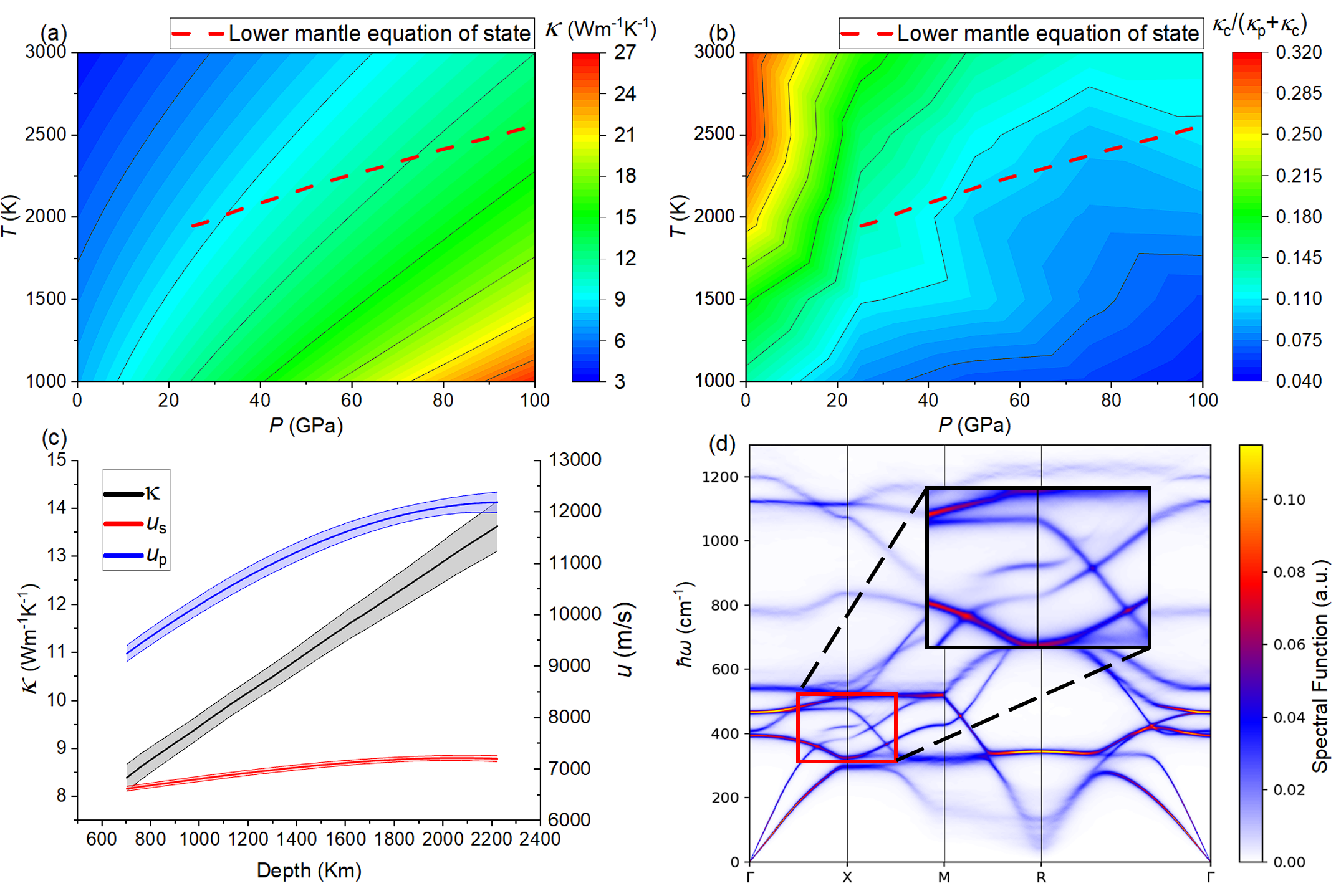}
\caption{(a) Thermal conductivity of \ch{CaSiO3} at different pressures and temperatures, (b) \(\frac{\kappa_{c}}{\kappa_{p}+\kappa_{c}}\), where \(\kappa_{p}\) and \(\kappa_{c}\) are particle-like and wave-like contributions to the thermal conductivity. The purple dashed line represents the lower mantle's equation of state. (c) Thermal conductivity $\kappa$, transverse sound velocity $u_s$ and longitudinal sound velocity $u_p$ as a function of depth. (d) Spectral function of \ch{CaSiO3} at \SI{100}{\giga\pascal} and \SI{2500}{\kelvin}. A small satellite band appears between $\SI{400}{\centi\meter}^{-1}$ and $\SI{500}{\centi\meter}^{-1}$ at symmetry point $X$ (highlighted in the red box), showcasing the strong anharmonicity of \ch{CaSiO3} even at very high pressure.
}
\label{fig:figure4_thermal_conductivity}
\end{figure*}

\subsection{Elastic properties and sound velocity}

Geological models can be validated by analyzing seismic wave propagation through the lower mantle, which requires knowledge of the elastic properties of the materials that compose it. Unfortunately, \ch{CaSiO3} is unquenchable at room conditions, and due to the difficulties in synthesizing samples sufficiently large to measure sound velocities, very few experimental data are available at high pressures. For this reason, first-principles simulations are crucial. However, recent works showed how state-of-the-art approaches fail to reproduce the experimental elastic properties at low pressures\cite{greaux2019sound}.
In particular, static DFT systematically overestimates the shear modulus\cite{Karki1998,Karki2001,stixrude2007phase} and, consequently, the transverse wave speed, while FPMD simulations improve in agreement with experiments\cite{kawai2015small}. 

Two hypotheses were suggested to explain the low shear modulus of \ch{CaSiO3}. The first assumes the stability of the tetragonal phase in the lower mantle\cite{stixrude2007phase, li2006phase}, which we already disproved in the first part of this work. The second hypothesis asserts that shear favors a rotation of oxygen octahedra similar to the tetragonal distortion, thus hypothesizing that the propagation of the transverse sound wave occurs similarly between the cubic and the tetragonal phase\cite{kawai2015small,caracas2005casio3}. 
However, if a shear deformation of the cubic cell were to result in a rotation of oxygen octahedra, then the cubic phase would become unstable at high temperatures under a finite shear strain. In particular, at least one of the triple degenerate low-frequency phonon modes at R of the free energy Hessian, associated with the rotation of the oxygen octahedra, would become imaginary. To test this hypothesis, we calculated the free energy Hessian of a shear strained 5-atom cubic structure at \SI{100}{\giga\pascal} and \SI{3000}{\kelvin} (\figurename~\ref{fig:figure3}c), and indeed the triple degenerate low-frequency mode at R becomes split by the shear (the presence of the shear introduces a favorite axis for the oxygen octahedral rotation). However, all three modes remain positive, with no sign of instability. Moreover, even the frequency of the lowest one increases under shear when compared to the cubic structure.
This indicates that shear increases the free energy barrier to rotate the oxygen octahedra even at 3000 K, disproving the hypothesis that shear stabilizes octahedral tilts at high temperatures.

We evaluate the isothermal sound velocity from the slope of the acoustic physical phonons at finite temperatures. For this purpose, we employ the phonons as obtained from the dispersion of the free energy Hessian, which coincides with the poles of the dynamical Green's function (the actual physical excitations of the lattice in a strongly anharmonic material)\cite{bianco2017second,monacelli2021time} in the $\omega\rightarrow0$ limit. This condition is realized in the acoustic modes near $\Gamma$. The free energy Hessian accounts for a negative-semidefinite ``bubble'' diagram\cite{bianco2017second,Siciliano2023} that reduces the slope of the acoustic phonons and is the real reason for the lower shear modulus at finite temperatures reported by FPMD compared to harmonic predictions\cite{kawai2015small}.
We reach a good agreement with experimental data for the longitudinal sound velocity (\figurename~\ref{fig:figure3}b); the simulated transverse sound velocity (\figurename~\ref{fig:figure3}a) agrees with predictions from FPMD and it improves the experimental agreement over static DFT calculations\cite{li2006phase}, but it is still significantly overestimated. 
This is due to the simulations' harder shear modulus value than experimental measurements. Therefore, the remaining possible explanation for the discrepancy is the limited precision of the exchange-correlation functional, here employed at the generalized-gradient approximation (GGA) within the PBEsol scheme\cite{PBEsol}.

\subsection{Thermal transport}
Mass transfer is impeded at the boundary separating the Earth's core from the lower mantle; thus, the mechanism driving heat transport predominantly relies on thermal diffusion.
Here, we report the lattice thermal conductivity of \ch{CaSiO3} at the thermodynamic conditions of the lower mantle and investigate the role played by the anharmonicity of the crystal at high temperatures. 
A previous study investigated thermal transport of \ch{CaSiO3}\cite{wentzcovitch2021thermal} employing the Boltzmann transport equation.
In \figurename~\ref{fig:figure4_thermal_conductivity}d, we report the phonon spectra, including physical linewidth at \SI{0}{\giga\pascal} and \SI{1000}{\kelvin}, computed through the time-dependent self-consistent harmonic approximation\cite{monacelli2021time,Siciliano2023}. The overlap between the phonon spectra in the 300-\SI{600}{\per\centi\meter} energy range highlights the possible tunneling between phonons at high temperatures. Moreover, while the phonon scattering rates are lower than their frequency, the phonon spectrum presents broad features and satellite peaks (\figurename~ \ref{fig:figure4_thermal_conductivity}d), thus questioning the applicability of the Boltzmann transport theory and the need to explore the role played by coherent transport across different phonon bands. This can be inverstigated through the Wigner formulation of thermal transport \cite{simoncelli2019unified,di2023crossover,klarbring2020anharmonicity, Caldarelli2022}, which accounts for both wave-like tunneling between different phonon modes and the standard particle-like (Peierls) propagation.
%

Since the anharmonic dressing of the phonons stabilizes the structure, providing positive-definite phonons, it allows us to employ the anharmonic renormalized phonon bands for calculating thermal transport in the cubic \ch{CaSiO3}, which would be impossible within the standard perturbation theory due to the imaginary harmonic phonon bands. In this way, we also account for the thermal effects that significantly harden M-R low-frequency bands and soften the high-energy breathing modes of oxygen. 

The present results show that the Wigner diffusion of the heat wave of \ch{CaSiO3} is suppressed by high pressure in the lower mantle. In contrast, it plays a major role at ambient pressure (\figurename~\ref{fig:figure4_thermal_conductivity}b). This can be understood in detail by looking at the distribution of phonon lifetimes (see SI. 3 in supplementary information): In the high pressure and low temperature regime (SI. 3d), these are mainly localized above the so-called Wigner limit, and so particle-like propagation dominates thermal transport \cite{di2023crossover}. On the other hand, in the low pressure and high temperature regime (SI. 3a) the phonon lifetimes cloud lies on the Wigner limit and so wave-like effects become relevant \cite{di2023crossover}.

In conclusion, our findings provide crucial insights into the thermodynamics of \ch{CaSiO3} under Earth's lower mantle conditions. We demonstrated that \ch{CaSiO3} consistently exists in its cubic phase, with possible metastable tetragonal distortions observed above the transition temperature due to the identified first-order kind of phase transition. Notably, quantum fluctuations remain significant in the thermodynamics of this material, even at high temperatures, due to the strong covalent bonds that stiffen under pressure. Our results align well with available X-ray data and measurements of the longitudinal sound velocity, although they slightly overestimate the transverse sound velocity. Lastly, we presented the thermal conductivity of this material as a function of depth, highlighting that, despite the substantial anharmonicity of the phonon bands, its behavior is effectively captured by Boltzmann transport theory.

\section*{Methods}
The anharmonicity of \ch{CaSiO3} is described through the stochastic self-consistent harmonic approximation (SSCHA)\cite{monacelli2021stochastic}. A trial Gaussian density matrix for the nuclei is optimized to minimize the quantum free energy of the lattice. The trial Gaussian density matrix is parametrized through the average positions of atoms (centroids) and quantum-thermal nuclear fluctuations (covariance matrix)\cite{monacelli2024simulating}. The Gaussian constraint is the least possible biased choice: it is the distribution maximizing the entropy where centroids and covariance matrix are the only two parameters. A one-to-one mapping exists between a harmonic Hamiltonian and the corresponding Gaussian density matrix. This mapping allows for the direct entropy calculation of the trial density matrix. The SSCHA evaluates the quantum free energy as:
\begin{equation}
F \ge F[\tilde\rho]=\left<K+V\right>_{\tilde\rho}-TS[\tilde\rho],
\label{eq:F}
\end{equation}
where $K$ is the kinetic energy, $S$ is the entropy (both known analytically for any Gaussian density matrix), and $V(\boldsymbol{R})$ is the interatomic potential energy landscape. The average of Eq.~\eqref{eq:F} is computed with a Monte Carlo algorithm 
on randomly displaced ionic configurations described elsewhere\cite{monacelli2021stochastic}. For each random configuration, $V(\boldsymbol{R})$ is evaluated as the ground state energy of density-functional theory (DFT) calculations within the Born-Oppenheimer approximation. The DFT calculations are solved in a plane-wave basis set as implemented in Quantum ESPRESSO\cite{Giannozzi2017} with the PBEsol\cite{PBEsol} approximation for the exchange-correlation functional and a combination of ultrasoft and Projected Augmented Wave pseudopotentials from the SSSP library (precision version 1.3)\cite{SSSP3}. The plane-wave basis set is truncated using a cutoff 
of $80$ Ry for wavefunction kinetic energy and $640$ Ry for the charge density. We employed supercell calculation with 40 atoms for all phases (cubic, tetragonal, and cubic with a shear), sampling the Brillouin zone with a $2\times2\times2$ Monkhorst-Pack mesh. 

To evaluate the volume and cell shape as a function of both temperature and pressure, the lattice vectors are optimized to minimize the Gibbs free energy
\begin{equation}
    G = F - P\Omega
\end{equation}
where $\Omega$ is the volume of the primitive cell, and $P$ is the target pressure\cite{monacelli2021stochastic}.
The SSCHA has already been applied successfully to a wide range of strongly anharmonic systems, as in high-pressure hydrogen and hydrides\cite{Errea2020,MonacelliH2023,Monacelli2020}, low dimensional materials\cite{Aseginolaza2024,Romanin2021}, and second-order phase transitions in perovskite structures similar to \ch{CaSiO3} with unstable harmonic phonons\cite{MonacelliCsSnI32023, Ranalli2023, Verdi2023}. 

The SSCHA allows us to rigorously compute the Free energy landscape as a function of the average nuclear position. Thanks to the Landau theory of phase transitions, this can be employed to study energy barriers between phases and second-order phase transitions by computing the curvature of the free energy landscape along the order parameter. 
The curvature change in the free energy landscape at the point where the order parameter is zero (the high-symmetry phase) identifies the critical point of a second-order phase transition.
The SSCHA can compute the critical point for the tetragonal to cubic by evaluating the free energy Hessian\cite{bianco2017second}:
\begin{equation}
    \frac{\partial F}{\partial R_a \partial R_b} = \Phi_{ab} + \sum_{cdeflm}{\overset{(3)}{\Phi}}_{acd}[1-\Lambda {\overset{(4)}{\Phi}}]^{-1}_{cdef}\Lambda_{eflm}{\overset{(3)}{\Phi}}_{lmb}
\end{equation}

The sound velocity is evaluated with the static linear response theory at finite temperature within the \emph{bubble} approximation\cite{bianco2017second}, using Fourier interpolations to evaluate the response at small q-points not commensurate with the original simulation cell\cite{Bianco2018}. This allows us to determine precise sound velocities without requiring large supercells in molecular dynamics simulations. 
Since the simulation keeps the temperature fixed, perturbation theory computes the isothermal sound velocity. The fast vibration of seismic waves compared to the equilibration time requires the calculation of adiabatic sound velocity, which can be obtained with standard thermodynamic relations employing the computed bulk modulus and thermal expansion rate evaluated from the variable cell relaxation. The conversion between isentropic sound velocity and isothermal sound velocity is expressed as follows:
\begin{equation}
\begin{split}
u_{s,S} &= u_{s,T}
\\
u_{p,S} &= \sqrt{{u_{p,T}}^2+\frac{B_S - B_T}{\rho}},
\end{split}
\end{equation}
where $u_s$ and $u_p$ represent transverse and longitudinal sound velocities, while $B$ stands for bulk modulus, $\rho$ is the density of the material, and $S$ and $T$ indexes indicate isentropic and isothermal conditions, respectively. The isentropic bulk modulus here is obtained by
\begin{equation}
\frac{1}{B_S} = \frac{1}{B_T} +\alpha_{V}\frac{\left(\frac{\partial S}{\partial P}\right)_{T}}{\left(\frac{\partial S}{\partial T}\right)_{P}},
\end{equation}
where $\alpha_V$ is the thermal expansion coefficient, and $\left(\frac{\partial S}{\partial P}\right)_{T}$ and $\left(\frac{\partial S}{\partial T}\right)_{P}$ are partial derivatives of entropy with respect to pressure and temperature.


The dynamical extension of SSCHA (time-dependent SSCHA)\cite{monacelli2021time,Siciliano2023,libbi2024atomistic,libbi2024ultrafast} allows us to obtain interacting phonon Green functions describing lattice excitations. From the imaginary part of the phononic Green functions, we estimate the spectrum of phonons reported in \figurename~\ref{fig:figure4_thermal_conductivity}d. From the phonon spectral function, we derive phonon-quasiparticles' lifetimes to calculate the lattice thermal conductivity in a temperature range that includes the characteristic temperatures of the lower mantle. This is expressed following the Wigner formalism as \cite{simoncelli2019unified,simoncelli2022wigner,di2023crossover}.
\begin{align}
\kappa^{\alpha\beta}_{tot} &= \kappa^{\alpha\beta}_{P,SMA}+\frac{1}{{\left(2\pi\right)}^3}\int_{BZ}\sum_{s\neq s'}\frac{{\omega(\boldsymbol{q})}_s+{\omega(\boldsymbol{q})}_{s'}}{4} \nonumber
\\
&\times\left[\frac{{C(\boldsymbol{q})}_{s}}{{\omega(\boldsymbol{q})}_{s}}+\frac{{C(\boldsymbol{q})}_{s'}}{{\omega(\boldsymbol{q})}_{s'}}\right]
{V^\alpha(\boldsymbol{q})}_{s,s'}{V^\beta(\boldsymbol{q})}_{s',s}  \nonumber
\\
&\times\frac{\frac{1}{2}[\Gamma(\boldsymbol{q})_s+\Gamma(\boldsymbol{q})_{s'}]}{[\omega(\boldsymbol{q})_{s'}-\omega(\boldsymbol{q})_{s}]^2+\frac{1}{4}[\Gamma(\boldsymbol{q})_s+\Gamma(\boldsymbol{q})_{s'}]^2}d^3q,
\end{align}
where $\kappa^{\alpha\beta}_{P,SMA}$ stands for the Peierls-Boltzmann (particle-like) conductivity, governed by phonon-phonon scattering within the single-mode relaxation time approximation (SMA). The additional term is the positive-definite tensor ($\kappa^{\alpha\beta}_{C}$) arising from the wave-tunneling between two non-degenerate bands ($s\neq s'$). For each wave vector $\boldsymbol{q}$, the $\omega(\boldsymbol{q})_s$ represents the frequency of the $s$-phonon mode. The specific heat $C(\boldsymbol{q}_s)$ is calculated with equilibrium Bose-Einstein distribution $\overline{N}(\boldsymbol{q})_s$ at temperature $T$, $V^{\alpha}(\boldsymbol{q})_{s,s'}$ and $V^{\beta}(\boldsymbol{q})_{s,s'}$ represents the cartesian components of the velocity operator whose diagonal elements ($s=s'$) give phonon group velocity. $\Gamma(\boldsymbol{q})_s$ are phonon linewidths. From the Lorentzian profile, it is clearly seen that when phonon linewidths are comparable to the interband energy spacing then wave-like heat condunction yields a non-negligible contribution to the overall thermal transport. On the contrary, when the opposite regime is verified thermal transport is mainly of particle-like nature. The calculation of thermal transport is performed within the Phono3py software\cite{phono3py}.

\clearpage


\begin{thebibliography}{10}

\bibitem{katsura2010adiabatic}
T.~Katsura, A.~Yoneda, D.~Yamazaki, T.~Yoshino, and E.~Ito, ``Adiabatic
  temperature profile in the mantle,'' {\em Physics of the Earth and Planetary
  Interiors}, vol.~183, no.~1-2, pp.~212--218, 2010.

\bibitem{dziewonski1981preliminary}
A.~M. Dziewonski and D.~L. Anderson, ``Preliminary reference earth model,''
  {\em Physics of the earth and planetary interiors}, vol.~25, no.~4,
  pp.~297--356, 1981.

\bibitem{gasparik1994experimental}
T.~Gasparik, K.~Wolf, and C.~M. Smith, ``Experimental determination of phase
  relations in the casio3 system from 8 to 15 gpa,'' {\em American
  Mineralogist}, vol.~79, no.~11-12, pp.~1219--1222, 1994.

\bibitem{shen1995measurement}
G.~Shen and P.~Lazor, ``Measurement of melting temperatures of some minerals
  under lower mantle pressures,'' {\em Journal of Geophysical Research: Solid
  Earth}, vol.~100, no.~B9, pp.~17699--17713, 1995.

\bibitem{zerr1997melting}
A.~Zerr, G.~Serghiou, and R.~Boehler, ``Melting of casio3 perovskite to 430
  kbar and first in-situ measurements of lower mantle eutectic temperatures,''
  {\em Geophysical research letters}, vol.~24, no.~8, pp.~909--912, 1997.

\bibitem{braithwaite2019melting}
J.~Braithwaite and L.~Stixrude, ``Melting of casio3 perovskite at high
  pressure,'' {\em Geophysical Research Letters}, vol.~46, no.~4,
  pp.~2037--2044, 2019.

\bibitem{stixrude2007phase}
L.~Stixrude, C.~Lithgow-Bertelloni, B.~Kiefer, and P.~Fumagalli, ``Phase
  stability and shear softening in ca si o 3 perovskite at high pressure,''
  {\em Physical Review B}, vol.~75, no.~2, p.~024108, 2007.

\bibitem{li2006phase}
L.~Li, D.~J. Weidner, J.~Brodholt, D.~Alfe, G.~D. Price, R.~Caracas, and
  R.~Wentzcovitch, ``Phase stability of casio3 perovskite at high pressure and
  temperature: Insights from ab initio molecular dynamics,'' {\em Physics of
  the Earth and Planetary Interiors}, vol.~155, no.~3-4, pp.~260--268, 2006.

\bibitem{adams2006ab}
D.~J. Adams and A.~R. Oganov, ``Ab initio molecular dynamics study of ca si o 3
  perovskite at p- t conditions of earth’s lower mantle,'' {\em Physical
  Review B}, vol.~73, no.~18, p.~184106, 2006.

\bibitem{li2006elasticity}
L.~Li, D.~J. Weidner, J.~Brodholt, D.~Alfe, G.~D. Price, R.~Caracas, and
  R.~Wentzcovitch, ``Elasticity of casio3 perovskite at high pressure and high
  temperature,'' {\em Physics of the Earth and Planetary Interiors}, vol.~155,
  no.~3-4, pp.~249--259, 2006.

\bibitem{thomson2019seismic}
A.~Thomson, W.~Crichton, J.~Brodholt, I.~Wood, N.~Siersch, J.~Muir, D.~Dobson,
  and S.~A. Hunt, ``Seismic velocities of casio3 perovskite can explain llsvps
  in earth’s lower mantle,'' {\em Nature}, vol.~572, no.~7771, pp.~643--647,
  2019.

\bibitem{sun2016confirming}
N.~Sun, Z.~Mao, S.~Yan, X.~Wu, V.~B. Prakapenka, and J.-F. Lin, ``Confirming a
  pyrolitic lower mantle using self-consistent pressure scales and new
  constraints on casio3 perovskite,'' {\em Journal of Geophysical Research:
  Solid Earth}, vol.~121, no.~7, pp.~4876--4894, 2016.

\bibitem{kurashina2004phase}
T.~Kurashina, K.~Hirose, S.~Ono, N.~Sata, and Y.~Ohishi, ``Phase transition in
  al-bearing casio3 perovskite: implications for seismic discontinuities in the
  lower mantle,'' {\em Physics of the Earth and Planetary Interiors}, vol.~145,
  no.~1-4, pp.~67--74, 2004.

\bibitem{ricolleau2010phase}
A.~Ricolleau, J.-p. Perrillat, G.~Fiquet, I.~Daniel, J.~Matas, A.~Addad,
  N.~Menguy, H.~Cardon, M.~Mezouar, and N.~Guignot, ``Phase relations and
  equation of state of a natural morb: Implications for the density profile of
  subducted oceanic crust in the earth's lower mantle,'' {\em Journal of
  Geophysical Research: Solid Earth}, vol.~115, no.~B8, 2010.

\bibitem{immoor2022weak}
J.~Immoor, L.~Miyagi, H.-P. Liermann, S.~Speziale, K.~Schulze, J.~Buchen,
  A.~Kurnosov, and H.~Marquardt, ``Weak cubic casio3 perovskite in the
  earth’s mantle,'' {\em Nature}, vol.~603, no.~7900, pp.~276--279, 2022.

\bibitem{miyagi2009diamond}
L.~Miyagi, S.~Merkel, T.~Yagi, N.~Sata, Y.~Ohishi, and H.-R. Wenk, ``Diamond
  anvil cell deformation of casio3 perovskite up to 49 gpa,'' {\em Physics of
  the Earth and Planetary Interiors}, vol.~174, no.~1-4, pp.~159--164, 2009.

\bibitem{shieh2004elasticity}
S.~R. Shieh, T.~S. Duffy, and G.~Shen, ``Elasticity and strength of calcium
  silicate perovskite at lower mantle pressures,'' {\em Physics of the Earth
  and Planetary Interiors}, vol.~143, pp.~93--105, 2004.

\bibitem{simoncelli2019unified}
M.~Simoncelli, N.~Marzari, and F.~Mauri, ``Unified theory of thermal transport
  in crystals and glasses,'' {\em Nature Physics}, vol.~15, no.~8,
  pp.~809--813, 2019.

\bibitem{phono3py}
A.~Togo, L.~Chaput, and I.~Tanaka, ``Distributions of phonon lifetimes in
  brillouin zones,'' {\em Phys. Rev. B}, vol.~91, p.~094306, Mar 2015.

\bibitem{ShengBTE_2014}
W.~Li, J.~Carrete, N.~A. Katcho, and N.~Mingo, ``{ShengBTE:} a solver of the
  {B}oltzmann transport equation for phonons,'' {\em Comp. Phys. Commun.},
  vol.~185, p.~1747–1758, 2014.

\bibitem{carrete2017almabte}
J.~Carrete, B.~Vermeersch, A.~Katre, A.~van Roekeghem, T.~Wang, G.~K. Madsen,
  and N.~Mingo, ``almabte: A solver of the space--time dependent boltzmann
  transport equation for phonons in structured materials,'' {\em Computer
  Physics Communications}, vol.~220, pp.~351--362, 2017.

\bibitem{raya2022bte}
M.~Raya-Moreno, X.~Cartoix{\`a}, and J.~Carrete, ``Bte-barna: An extension of
  almabte for thermal simulation of devices based on 2d materials,'' {\em
  Computer Physics Communications}, vol.~281, p.~108504, 2022.

\bibitem{ding2022observation}
Z.~Ding, K.~Chen, B.~Song, J.~Shin, A.~A. Maznev, K.~A. Nelson, and G.~Chen,
  ``Observation of second sound in graphite over 200 k,'' {\em Nature
  communications}, vol.~13, no.~1, p.~285, 2022.

\bibitem{huberman2019observation}
S.~Huberman, R.~A. Duncan, K.~Chen, B.~Song, V.~Chiloyan, Z.~Ding, A.~A.
  Maznev, G.~Chen, and K.~A. Nelson, ``Observation of second sound in graphite
  at temperatures above 100 k,'' {\em Science}, vol.~364, no.~6438,
  pp.~375--379, 2019.

\bibitem{huang2023observation}
X.~Huang, Y.~Guo, Y.~Wu, S.~Masubuchi, K.~Watanabe, T.~Taniguchi, Z.~Zhang,
  S.~Volz, T.~Machida, and M.~Nomura, ``Observation of phonon poiseuille flow
  in isotopically purified graphite ribbons,'' {\em Nature Communications},
  vol.~14, no.~1, p.~2044, 2023.

\bibitem{lee2015hydrodynamic}
S.~Lee, D.~Broido, K.~Esfarjani, and G.~Chen, ``Hydrodynamic phonon transport
  in suspended graphene,'' {\em Nature communications}, vol.~6, no.~1, p.~6290,
  2015.

\bibitem{cepellotti2017boltzmann}
A.~Cepellotti and N.~Marzari, ``Boltzmann transport in nanostructures as a
  friction effect,'' {\em Nano letters}, vol.~17, no.~8, pp.~4675--4682, 2017.

\bibitem{sendra2022hydrodynamic}
L.~Sendra, A.~Beardo, J.~Bafaluy, P.~Torres, F.~X. Alvarez, and J.~Camacho,
  ``Hydrodynamic heat transport in dielectric crystals in the collective limit
  and the drifting/driftless velocity conundrum,'' {\em Physical Review B},
  vol.~106, no.~15, p.~155301, 2022.

\bibitem{jeong2021transient}
J.~Jeong, X.~Li, S.~Lee, L.~Shi, and Y.~Wang, ``Transient hydrodynamic lattice
  cooling by picosecond laser irradiation of graphite,'' {\em Physical Review
  Letters}, vol.~127, no.~8, p.~085901, 2021.

\bibitem{restuccia2023non}
L.~Restuccia and D.~Jou, ``Non-local vectorial internal variables and
  generalized guyer-krumhansl evolution equations for the heat flux,'' {\em
  Entropy}, vol.~25, no.~9, p.~1259, 2023.

\bibitem{sykora2023multiscale}
M.~S{\`y}kora, M.~Pavelka, L.~Restuccia, and D.~Jou, ``Multiscale heat
  transport with inertia and thermal vortices,'' {\em Physica Scripta},
  vol.~98, no.~10, p.~105234, 2023.

\bibitem{di2024vorticity}
E.~Di~Lucente, F.~Libbi, and N.~Marzari, ``Vorticity and compressibility
  hydrodynamics in electron and phonon fluids,'' {\em Bulletin of the American
  Physical Society}, 2024.

\bibitem{ponet2024energy}
L.~Ponet, E.~Di~Lucente, and N.~Marzari, ``The energy landscape of magnetic
  materials,'' {\em npj Computational Materials}, vol.~10, no.~1, p.~151, 2024.

\bibitem{Lay2008}
T.~Lay, J.~Hernlund, and B.~A. Buffett, ``Core{\textendash}mantle boundary heat
  flow,'' {\em Nature Geoscience}, vol.~1, pp.~25--32, Jan. 2008.

\bibitem{alfe1999melting}
D.~Alfe, M.~Gillan, and G.~Price, ``The melting curve of iron at the pressures
  of the earth's core from ab initio calculations,'' {\em Nature}, vol.~401,
  no.~6752, pp.~462--464, 1999.

\bibitem{sagatova2021phase}
D.~Sagatova, A.~Shatskiy, N.~Sagatov, and K.~Litasov, ``Phase relations in
  casio 3 system up to 100 gpa and 2500 k,'' {\em Geochemistry International},
  vol.~59, pp.~791--800, 2021.

\bibitem{wentzcovitch1995ab}
R.~M. Wentzcovitch, N.~L. Ross, and G.~Price, ``Ab initio study of mgsio3 and
  casio3 perovskites at lower-mantle pressures,'' {\em Physics of the Earth and
  Planetary Interiors}, vol.~90, no.~1-2, pp.~101--112, 1995.

\bibitem{caracas2006theoretical}
R.~Caracas and R.~M. Wentzcovitch, ``Theoretical determination of the
  structures of casio3 perovskites,'' {\em Acta Crystallographica Section B:
  Structural Science}, vol.~62, no.~6, pp.~1025--1030, 2006.

\bibitem{karki2001high}
B.~B. Karki, L.~Stixrude, and R.~M. Wentzcovitch, ``High-pressure elastic
  properties of major materials of earth's mantle from first principles,'' {\em
  Reviews of Geophysics}, vol.~39, no.~4, pp.~507--534, 2001.

\bibitem{sokolova2021equations}
T.~S. Sokolova and P.~I. Dorogokupets, ``Equations of state of ca-silicates and
  phase diagram of the casio3 system under upper mantle conditions,'' {\em
  Minerals}, vol.~11, no.~3, p.~322, 2021.

\bibitem{sun2014dynamic}
T.~Sun, D.-B. Zhang, and R.~M. Wentzcovitch, ``Dynamic stabilization of cubic
  ca si o 3 perovskite at high temperatures and pressures from ab initio
  molecular dynamics,'' {\em Physical Review B}, vol.~89, no.~9, p.~094109,
  2014.

\bibitem{wu2024deep}
F.~Wu, Y.~Sun, T.~Wan, S.~Wu, and R.~M. Wentzcovitch, ``Deep-learning-based
  prediction of the tetragonal→ cubic transition in davemaoite,'' {\em
  Geophysical Research Letters}, vol.~51, no.~12, p.~e2023GL108012, 2024.

\bibitem{greaux2019sound}
S.~Gr{\'e}aux, T.~Irifune, Y.~Higo, Y.~Tange, T.~Arimoto, Z.~Liu, and
  A.~Yamada, ``Sound velocity of casio3 perovskite suggests the presence of
  basaltic crust in the earth’s lower mantle,'' {\em Nature}, vol.~565,
  no.~7738, pp.~218--221, 2019.

\bibitem{hofmeister2019heat}
A.~Hofmeister, {\em Heat transport and energetics of the earth and rocky
  planets}.
\newblock Elsevier, 2019.

\bibitem{lodders1998planetary}
K.~Lodders and B.~Fegley, {\em The planetary scientist's companion}.
\newblock Oxford University Press on Demand, 1998.

\bibitem{schubert2015treatise}
G.~Schubert, {\em Treatise on geophysics}.
\newblock Elsevier, 2015.

\bibitem{rost2013core}
S.~Rost, ``Core--mantle boundary landscapes,'' {\em Nature Geoscience}, vol.~6,
  no.~2, pp.~89--90, 2013.

\bibitem{wentzcovitch2021thermal}
Z.~Zhang, D.-B. Zhang, K.~Onga, A.~Hasegawa, K.~Ohta, K.~Hirose, and R.~M.
  Wentzcovitch, ``Thermal conductivity of casio 3 perovskite at lower mantle
  conditions,'' {\em Physical Review B}, vol.~104, no.~18, p.~184101, 2021.

\bibitem{di2023crossover}
E.~Di~Lucente, M.~Simoncelli, and N.~Marzari, ``Crossover from boltzmann to
  wigner thermal transport in thermoelectric skutterudites,'' {\em Phys. Rev.
  Res.}, vol.~5, p.~033125, Aug 2023.

\bibitem{PBEsol}
J.~P. Perdew, A.~Ruzsinszky, G.~I. Csonka, O.~A. Vydrov, G.~E. Scuseria, L.~A.
  Constantin, X.~Zhou, and K.~Burke, ``Restoring the density-gradient expansion
  for exchange in solids and surfaces,'' {\em Physical Review Letters},
  vol.~100, no.~13, p.~136406.

\bibitem{Errea2014a}
I.~Errea, M.~Calandra, and F.~Mauri, ``Anharmonic free energies and phonon
  dispersions from the stochastic self-consistent harmonic approximation:
  Application to platinum and palladium hydrides,'' {\em Physical Review B},
  vol.~89, no.~6, p.~064302, 2014.

\bibitem{bianco2017second}
R.~Bianco, I.~Errea, L.~Paulatto, M.~Calandra, and F.~Mauri, ``Second-order
  structural phase transitions, free energy curvature, and
  temperature-dependent anharmonic phonons in the self-consistent harmonic
  approximation: Theory and stochastic implementation,'' {\em Physical Review
  B}, vol.~96, no.~1, p.~014111, 2017.

\bibitem{Monacelli2018}
L.~Monacelli, I.~Errea, M.~Calandra, and F.~Mauri, ``Pressure and stress tensor
  of complex anharmonic crystals within the stochastic self-consistent harmonic
  approximation,'' {\em Phys. Rev. B}, vol.~98, p.~024106, Jul 2018.

\bibitem{monacelli2021stochastic}
L.~Monacelli, R.~Bianco, M.~Cherubini, M.~Calandra, I.~Errea, and F.~Mauri,
  ``The stochastic self-consistent harmonic approximation: calculating
  vibrational properties of materials with full quantum and anharmonic
  effects,'' {\em Journal of Physics: Condensed Matter}, vol.~33, no.~36,
  p.~363001, 2021.

\bibitem{monacelli2024simulating}
L.~Monacelli, ``Simulating anharmonic crystals: Lights and shadows of
  first-principles approaches,'' {\em arXiv preprint arXiv:2407.03090}, 2024.

\bibitem{miotto2024fast}
M.~Miotto and L.~Monacelli, ``Fast prediction of anharmonic vibrational spectra
  for complex organic molecules,'' {\em npj Computational Materials}, vol.~10,
  Oct. 2024.

\bibitem{pasciak2018dynamic}
M.~Pa{\'s}ciak, T.~Welberry, J.~Kulda, S.~Leoni, and J.~Hlinka, ``Dynamic
  displacement disorder of cubic batio 3,'' {\em Physical review letters},
  vol.~120, no.~16, p.~167601, 2018.

\bibitem{Gigli2022}
L.~Gigli, M.~Veit, M.~Kotiuga, G.~Pizzi, N.~Marzari, and M.~Ceriotti,
  ``Thermodynamics and dielectric response of {BaTiO}3 by data-driven
  modeling,'' {\em npj Computational Materials}, vol.~8, p.~209, Sept. 2022.

\bibitem{MonacelliCsSnI32023}
L.~Monacelli and N.~Marzari, ``First-principles thermodynamics of
  {CsSnI}$_3$,'' {\em Chemistry of Materials}, vol.~35, pp.~1702--1709, Feb.
  2023.

\bibitem{kotiuga2022microscopic}
M.~Kotiuga, S.~Halilov, B.~Kozinsky, M.~Fornari, N.~Marzari, and G.~Pizzi,
  ``Microscopic picture of paraelectric perovskites from structural
  prototypes,'' {\em Physical Review Research}, vol.~4, no.~1, p.~L012042,
  2022.

\bibitem{ono2004phase}
S.~Ono, Y.~Ohishi, and K.~Mibe, ``Phase transition of ca-perovskite and
  stability of al-bearing mg-perovskite in the lower mantle,'' {\em American
  Mineralogist}, vol.~89, no.~10, pp.~1480--1485, 2004.

\bibitem{komabayashi2007phase}
T.~Komabayashi, K.~Hirose, N.~Sata, Y.~Ohishi, and L.~S. Dubrovinsky, ``Phase
  transition in casio3 perovskite,'' {\em Earth and Planetary Science Letters},
  vol.~260, no.~3-4, pp.~564--569, 2007.

\bibitem{chen2018crystal}
H.~Chen, S.-H. Shim, K.~Leinenweber, V.~Prakapenka, Y.~Meng, and C.~Prescher,
  ``Crystal structure of casio3 perovskite at 28--62 gpa and 300 k under
  quasi-hydrostatic stress conditions,'' {\em American Mineralogist: Journal of
  Earth and Planetary Materials}, vol.~103, no.~3, pp.~462--468, 2018.

\bibitem{Karki1998}
B.~B. Karki and J.~Crain, ``Structure and elasticity of {CaO} at high
  pressure,'' vol.~103, pp.~12405--12411.

\bibitem{Karki2001}
B.~B. Karki, L.~Stixrude, and R.~M. Wentzcovitch, ``High-pressure elastic
  properties of major materials of earth's mantle from first principles,''
  vol.~39, no.~4, pp.~507--534.

\bibitem{kawai2015small}
K.~Kawai and T.~Tsuchiya, ``Small shear modulus of cubic casio3 perovskite,''
  {\em Geophysical Research Letters}, vol.~42, no.~8, pp.~2718--2726, 2015.

\bibitem{caracas2005casio3}
R.~Caracas, R.~Wentzcovitch, G.~D. Price, and J.~Brodholt, ``Casio3 perovskite
  at lower mantle pressures,'' {\em Geophysical Research Letters}, vol.~32,
  no.~6, 2005.

\bibitem{monacelli2021time}
L.~Monacelli and F.~Mauri, ``Time-dependent self-consistent harmonic
  approximation: Anharmonic nuclear quantum dynamics and time correlation
  functions,'' {\em Physical Review B}, vol.~103, no.~10, p.~104305, 2021.

\bibitem{Siciliano2023}
A.~Siciliano, L.~Monacelli, G.~Caldarelli, and F.~Mauri, ``Wigner gaussian
  dynamics: Simulating the anharmonic and quantum ionic motion,'' {\em Phys.
  Rev. B}, vol.~107, p.~174307, May 2023.

\bibitem{klarbring2020anharmonicity}
J.~Klarbring, O.~Hellman, I.~A. Abrikosov, and S.~I. Simak, ``Anharmonicity and
  ultralow thermal conductivity in lead-free halide double perovskites,'' {\em
  Physical Review Letters}, vol.~125, no.~4, p.~045701, 2020.

\bibitem{Caldarelli2022}
G.~Caldarelli, M.~Simoncelli, N.~Marzari, F.~Mauri, and L.~Benfatto,
  ``Many-body green's function approach to lattice thermal transport,'' {\em
  Phys. Rev. B}, vol.~106, p.~024312, Jul 2022.

\bibitem{Giannozzi2017}
P.~Giannozzi, O.~Andreussi, T.~Brumme, O.~Bunau, M.~Buongiorno~Nardelli,
  M.~Calandra, R.~Car, C.~Cavazzoni, D.~Ceresoli, M.~Cococcioni, N.~Colonna,
  I.~Carnimeo, A.~Dal~Corso, S.~de~Gironcoli, P.~Delugas, R.~A. DiStasio,
  A.~Ferretti, A.~Floris, G.~Fratesi, G.~Fugallo, R.~Gebauer, U.~Gerstmann,
  F.~Giustino, T.~Gorni, J.~Jia, M.~Kawamura, H.-Y. Ko, A.~Kokalj,
  E.~Küçükbenli, M.~Lazzeri, M.~Marsili, N.~Marzari, F.~Mauri, N.~L. Nguyen,
  H.-V. Nguyen, A.~Otero-de-la Roza, L.~Paulatto, S.~Poncé, D.~Rocca,
  R.~Sabatini, B.~Santra, M.~Schlipf, A.~P. Seitsonen, A.~Smogunov, I.~Timrov,
  T.~Thonhauser, P.~Umari, N.~Vast, X.~Wu, and S.~Baroni, ``Advanced
  capabilities for materials modelling with quantum espresso,'' {\em Journal of
  Physics: Condensed Matter}, vol.~29, p.~465901, Oct. 2017.

\bibitem{SSSP3}
K.~Lejaeghere, G.~Bihlmayer, T.~Bj{\"o}rkman, P.~Blaha, S.~Bl{\"u}gel, V.~Blum,
  D.~Caliste, I.~E. Castelli, S.~J. Clark, A.~Dal~Corso, {\em et~al.},
  ``Reproducibility in density functional theory calculations of solids,'' {\em
  Science}, vol.~351, no.~6280, p.~1415, 2016.

\bibitem{Errea2020}
I.~Errea, F.~Belli, L.~Monacelli, A.~Sanna, T.~Koretsune, T.~Tadano, R.~Bianco,
  M.~Calandra, R.~Arita, F.~Mauri, and J.~A. Flores-Livas, ``Quantum crystal
  structure in the 250-kelvin superconducting lanthanum hydride,'' {\em
  Nature}, vol.~578, pp.~66--69, Feb. 2020.

\bibitem{MonacelliH2023}
L.~Monacelli, M.~Casula, K.~Nakano, S.~Sorella, and F.~Mauri, ``Quantum phase
  diagram of high-pressure hydrogen,'' {\em Nature Physics}, p.~845, Mar. 2023.

\bibitem{Monacelli2020}
L.~Monacelli, I.~Errea, M.~Calandra, and F.~Mauri, ``Black metal hydrogen above
  360{\hspace{0.167em}}{GPa} driven by proton quantum fluctuations,'' {\em
  Nature Physics}, vol.~17, pp.~63--67, Sept. 2020.

\bibitem{Aseginolaza2024}
U.~Aseginolaza, J.~Diego, T.~Cea, R.~Bianco, L.~Monacelli, F.~Libbi,
  M.~Calandra, A.~Bergara, F.~Mauri, and I.~Errea, ``Bending rigidity, sound
  propagation and ripples in flat graphene,'' {\em Nature Physics}, vol.~20,
  p.~1288–1293, May 2024.

\bibitem{Romanin2021}
D.~Romanin, L.~Monacelli, R.~Bianco, I.~Errea, F.~Mauri, and M.~Calandra,
  ``Dominant role of quantum anharmonicity in the stability and optical
  properties of infinite linear acetylenic carbon chains,'' {\em The Journal of
  Physical Chemistry Letters}, vol.~12, p.~10339–10345, Oct. 2021.

\bibitem{Ranalli2023}
L.~Ranalli, C.~Verdi, L.~Monacelli, G.~Kresse, M.~Calandra, and C.~Franchini,
  ``Temperature-dependent anharmonic phonons in quantum paraelectric {KTaO}$_3$
  by first principles and machine-learned force fields,'' {\em Advanced Quantum
  Technologies}, vol.~6, p.~2200131, Feb. 2023.

\bibitem{Verdi2023}
C.~Verdi, L.~Ranalli, C.~Franchini, and G.~Kresse, ``Quantum paraelectricity
  and structural phase transitions in strontium titanate beyond density
  functional theory,'' {\em Physical Review Materials}, vol.~7, Mar. 2023.

\bibitem{Bianco2018}
R.~Bianco, I.~Errea, M.~Calandra, and F.~Mauri, ``High-pressure phase diagram
  of hydrogen and deuterium sulfides from first principles: Structural and
  vibrational properties including quantum and anharmonic effects,'' {\em Phys.
  Rev. B}, vol.~97, p.~214101, Jun 2018.

\bibitem{libbi2024atomistic}
F.~Libbi, A.~Johansson, L.~Monacelli, and B.~Kozinsky, ``Atomistic simulations
  of out-of-equilibrium quantum nuclear dynamics,'' {\em arXiv preprint
  arXiv:2408.00902}, 2024.

\bibitem{libbi2024ultrafast}
F.~Libbi, A.~Johansson, B.~Kozinsky, and L.~Monacelli, ``Ultrafast quantum
  dynamics in srtio3 under impulsive thz radiation,'' {\em arXiv preprint
  arXiv:2408.12421}, 2024.

\bibitem{simoncelli2022wigner}
M.~Simoncelli, N.~Marzari, and F.~Mauri, ``Wigner formulation of thermal
  transport in solids,'' {\em Physical Review X}, vol.~12, no.~4, p.~041011,
  2022.

\end{thebibliography}


\end{document}


\pagestyle{fancy}
\lhead{\textit{Supplementary information}}

\title{
The thermodynamics of  CaSiO\textsubscript{3} in Earth's lower mantle: supplementary information}

\author{Yongjoong Shin}
\affiliation{Theory and Simulation of Materials (THEOS) and National Centre for Computational Design and Discovery of Novel Materials (MARVEL), École Polytechnique Fédérale de Lausanne, Lausanne 1015, Switzerland}

\author{Enrico Di Lucente}
\affiliation{Theory and Simulation of Materials (THEOS) and National Centre for Computational Design and Discovery of Novel Materials (MARVEL), École Polytechnique Fédérale de Lausanne, Lausanne 1015, Switzerland}

\author{Nicola Marzari}
\affiliation{Theory and Simulation of Materials (THEOS) and National Centre for Computational Design and Discovery of Novel Materials (MARVEL), École Polytechnique Fédérale de Lausanne, Lausanne 1015, Switzerland}
\affiliation{Laboratory for Materials Simulations, Paul Scherrer Institut, 5232 Villigen PSI, Switzerland}

\author{Lorenzo Monacelli}
\affiliation{Dipartimento di Fisica, Universitá di Roma La Sapienza, 00185 Roma, Italy}
\affiliation{Theory and Simulation of Materials (THEOS) and National Centre for Computational Design and Discovery of Novel Materials (MARVEL), École Polytechnique Fédérale de Lausanne, Lausanne 1015, Switzerland}



\maketitle

\begin{figure}[ht]
\centering
\includegraphics[width=0.7\columnwidth]{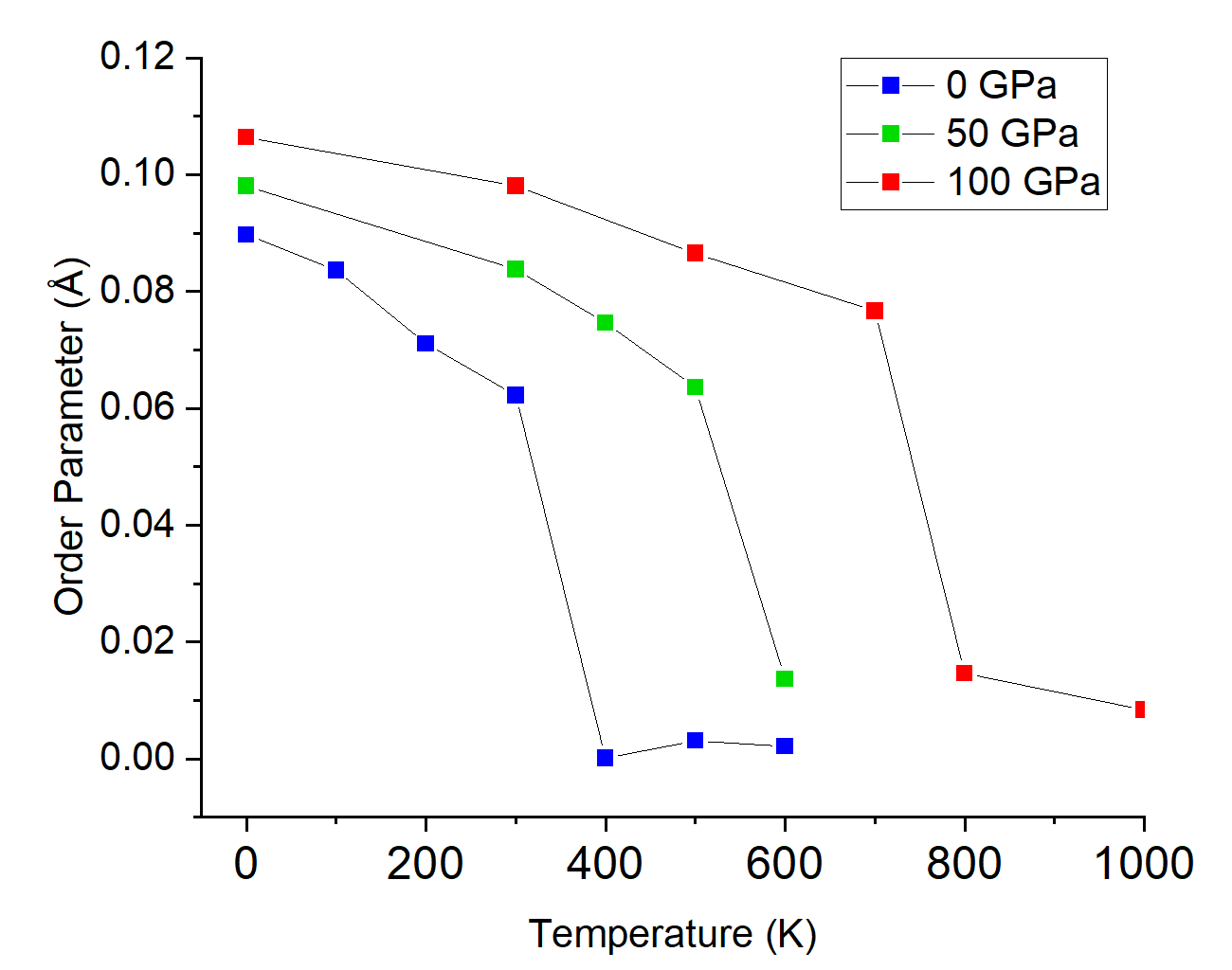}
\caption{Order parameter as a function of temperature at 0, 50, 100 GPa}
\label{fig:figure7}
\end{figure}

\begin{figure}[ht]
\centering
\includegraphics[width=0.7\columnwidth]{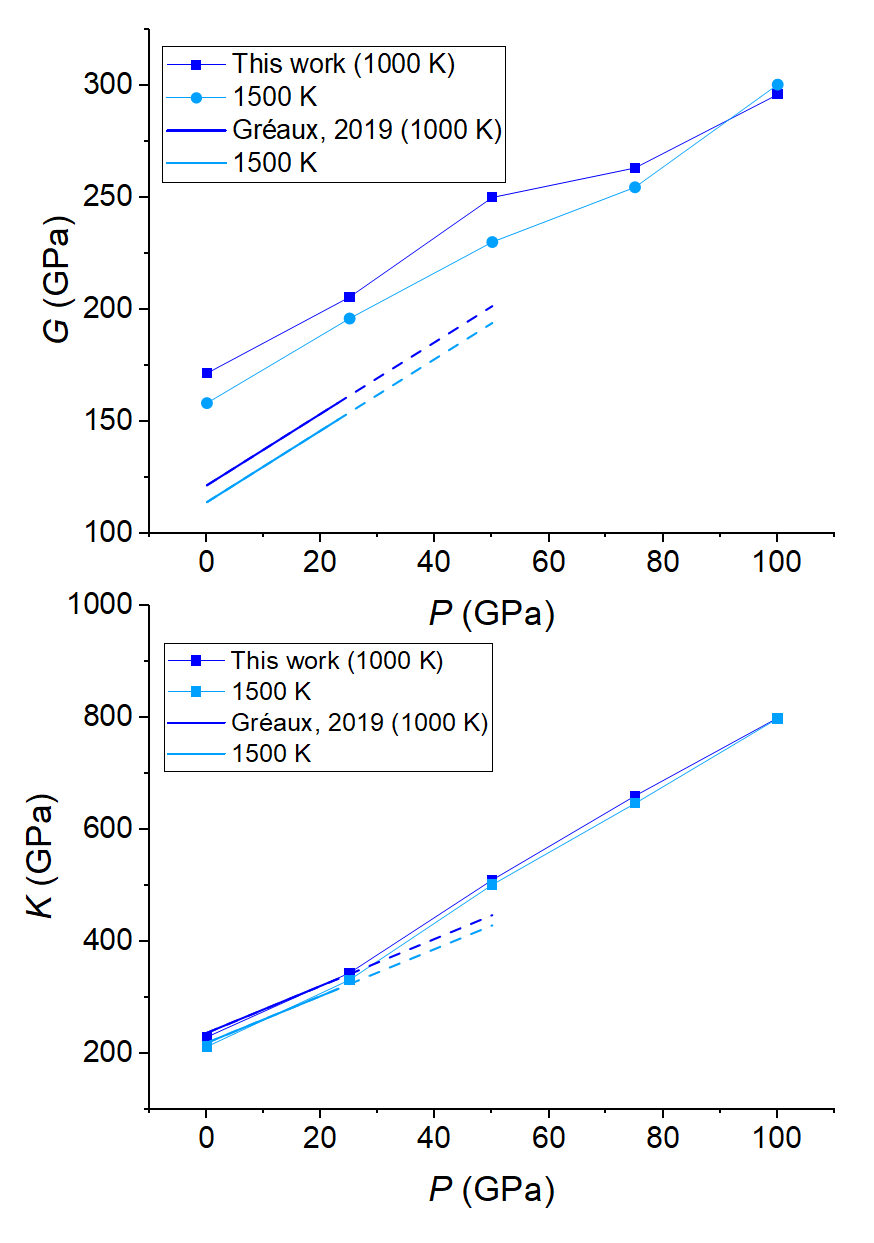}
\caption{(a) Shear modulus G and (b) bulk modulus K obtained from SSCHA (this work) and shear and bulk modulus-pressure relation estimated from experiment results\cite{greaux2019sound}. Comparing our results of shear modulus with experimental results, we find higher estimates than actual values. At \SI{25}{\giga\pascal} and \SI{1000}{\kelvin}, the shear modulus from experiments is \SI{25}{\giga\pascal}, which is about 26\% less then our predictions.}
\label{fig:figure7}
\end{figure}

\begin{figure*}[ht]
\centering
\includegraphics[width=\textwidth]{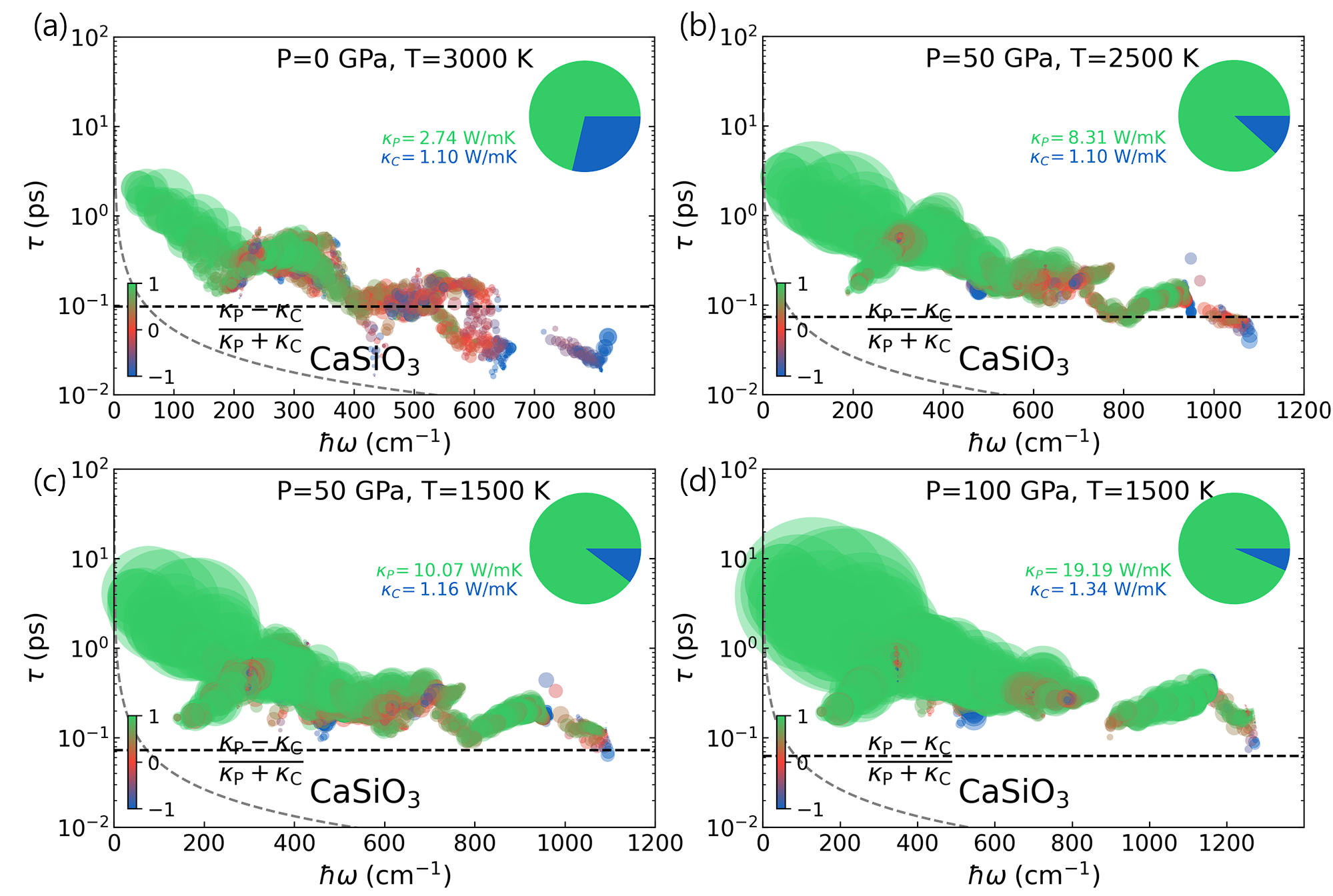}
\caption{Distribution of phonon lifetimes as a function of energy for different temperatures and pressures. The Wigner limit in time (dashed horizontal line) corresponds to a phonon lifetime equal to the inverse of the average interband spacing ($\tau^{\omega}=\Delta \omega_{avg}^{-1}$), and the dashed hyperbola shows the Ioffe-Regel limit in time($\tau^{IR}=\omega^{-1}$). The area of each scatter point is proportional to the contribution to the lattice thermal conductivity, and its color represent the origin of the contribution: $c = [\kappa_{p}-\kappa_{c}]/[\kappa_{p}+\kappa_{c}]$, where particle like is green and wavelike is blue.}
\label{fig:figure8}
\end{figure*}

\clearpage

\bibliographystyle{ieeetr}
\bibliography{bibliography.bib}